\documentclass[superscriptaddress,aps,jcp,twocolumn,nofootinbib,babel]{revtex4-1}
\usepackage{amsmath}
\usepackage{amssymb}
\usepackage{siunitx}
\usepackage{physics}
\usepackage{verbatim}
\usepackage{graphicx}
\newcommand\numberthis{\addtocounter{equation}{1}\tag{\theequation}}
\usepackage{times}	
\usepackage{tikz}
\newcommand{\ii}{{i}}
\newcommand*\circled[1]{\tikz[baseline=(char.base)]{
		\node[shape=circle,draw,inner sep=.5pt] (char) {#1};}}
\graphicspath{ {figure/} }
\usepackage{qcircuit}
\begin{document}

\title{Bravyi-Kitaev Superfast simulation of electronic structure on a quantum computer }

\author{Kanav Setia}
\affiliation{Department of Physics and Astronomy, Dartmouth College, Hanover, NH 03755}

\author{James D. Whitfield} 
\affiliation{Department of Physics and Astronomy, Dartmouth College, Hanover, NH 03755}
\email{james.d.whitfield@dartmouth.edu}
\date{\today}
\begin{abstract}
Present quantum computers often work with distinguishable qubits as their computational units.  In order to simulate indistinguishable fermionic particles, it is first required to map the fermionic state to the state of the qubits.  The Bravyi-Kitaev Superfast (BKSF) algorithm can be used to accomplish this mapping. The BKSF mapping has connections to quantum error correction and opens the door to new ways of understanding fermionic simulation in a topological context.  Here, we present the first detailed exposition of BKSF algorithm for molecular simulation. We provide the BKSF transformed qubit operators and report on our implementation of the BKSF fermion-to-qubits transform in OpenFermion.  In this initial study of the hydrogen molecule, we have compared BKSF, Jordan-Wigner and Bravyi-Kitaev transforms under the Trotter approximation. The gate count to implement BKSF is lower than Jordan-Wigner but higher than Bravyi-Kitaev. We considered different orderings of the exponentiated terms and found lower Trotter errors than previously reported for Jordan-Wigner and Bravyi-Kitaev algorithms. These results open the door to further study of the BKSF algorithm for quantum simulation.
\end{abstract}

\maketitle

\section{Introduction}

The development of the technology to build quantum computers has picked up pace over the last decade and is nearing a stage where the quantum computers can be used for commercial purposes \cite{Mohseni2017}. Groups around the world have been pursuing the construction of quantum computing devices in a wide variety of architectures such as superconducting quantum circuits \cite{Omalley2016,Barends2014,Kelly2014}, photons \cite{Peruzzo2014}, ion traps \cite{Munroe2016} and NMR \cite{Jiangfeng2010}. One major application area for these devices is simulating quantum mechanics.  This has spurred research into quantum algorithms for quantum simulation including quantum field theories \cite{Jordan2012}, lattice gauge theories \cite{Zohar2017}, and Markovian dynamics \cite{Childs2016}.  The present paper focuses on the quantum simulation of fermions which has received widespread attention in recent years \cite{Lanyon2010,Yung2014,Romero2017,Mcclean2017,Kandala2017,Bravyi2017}.

In general, qubits are distinguishable and individually addressable while fermions are not.  This conflict leads to a variety of strategies for the quantum simulation of fermions, including the Jordan-Wigner transformation \cite{Ortiz2001,Jordan1928}, the Bravyi-Kitaev transformation \cite{Havlicek2017,Seeley2012,Bravyi2000}, auxiliary fermion methods \cite{Ball2005,Verstraete2005,Havlicek2017} and LDPC codes \cite{Bravyi2017}. Other techniques were also proposed with explicit anti-symmetrization 
 \cite{Zalka1998,Lidar1999,Wiesner1996,Kassal2008,Lloyd1996}. In this paper we will focus on the Bravyi-Kitaev Superfast (BKSF) method \cite{Bravyi2000}.

  \begin{figure}
 	\includegraphics[width=8cm]{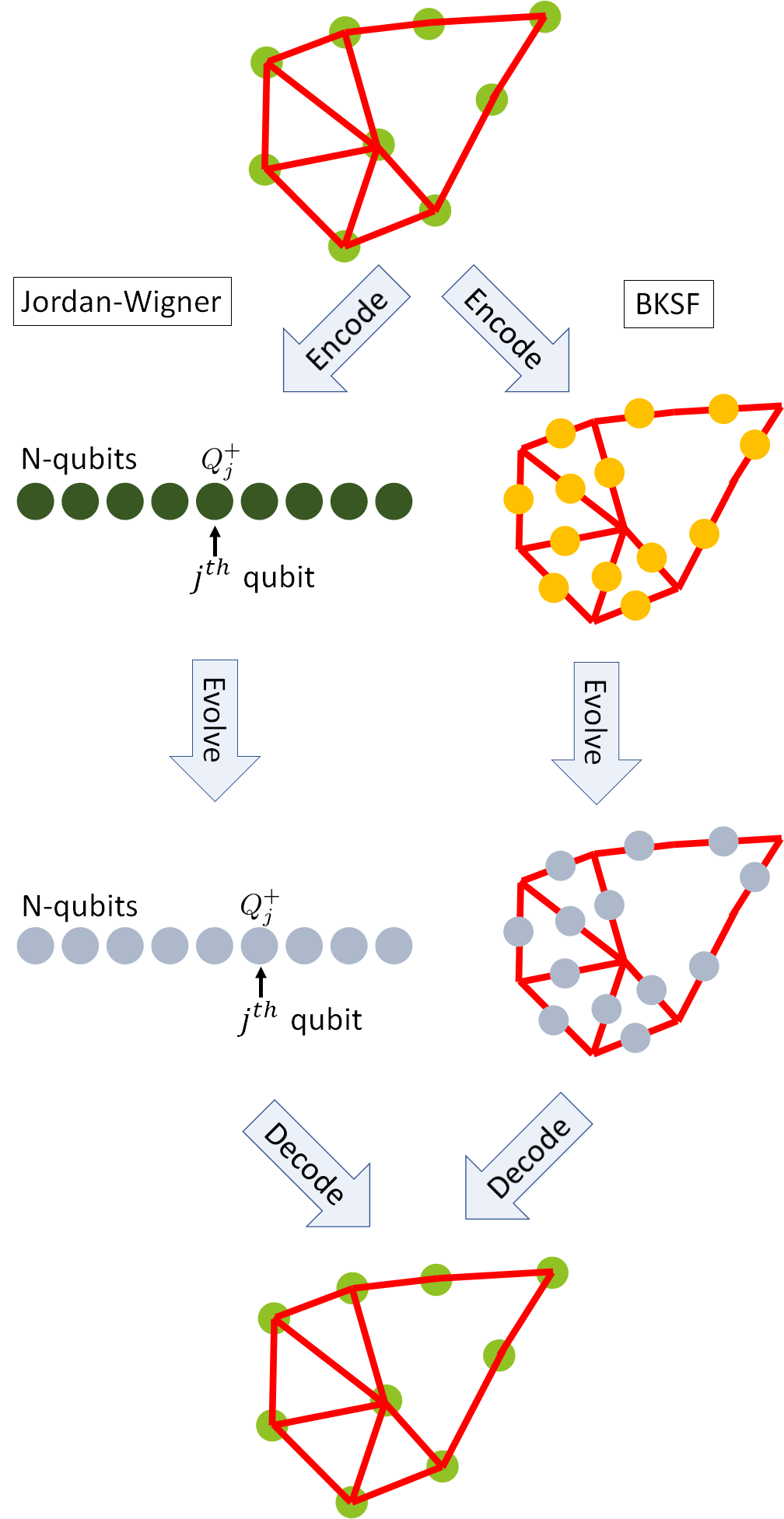}
 	\caption{Quantum simulation of fermionic systems requires encoding the fermionic system onto  qubits, studying their evolution over time and then decoding the qubits to get relevant fermionic system parameters. The Jordan-Wigner transformation, shown on the left, maps each mode to a qubit. The Bravyi-Kitaev Superfast (BKSF) transformation, shown on the right, maps each edge to a qubit. The schematic representation of the Bravyi-Kitaev transform is the same as the Jordan-Wigner. }
    \label{fig:pic_rep}
 \end{figure} 
 
Unlike other fermion-to-spin mappings, the fermionic operators in the BKSF algorithm do not explicitly depend on the number of modes.  Instead, the number of qubits required and the tensor locality of the fermionic operators depends on the specific interaction graph determined by the Hamiltonian. Figure \ref{fig:pic_rep} shows a pictorial representation of the BKSF transform in comparison to the Jordan-Wigner transform. For the Jordan-Wigner transform, each qubit corresponds to an orbital, while the BKSF transform uses qubits to represent the interaction terms between two orbitals.  In our previous work concerning the Hubbard model \cite{Havlicek2017}, it was shown that the BKSF algorithm has a clear advantages over the Jordan-Wigner and standard Bravyi-Kitaev algorithms.

In this article, we turn toward applications in molecular simulation starting with the hydrogen molecule. The purpose of this paper is to present the methodology of the BKSF algorithm and describe its implementation in OpenFermion \cite{Mcclean2017}. Further, the formalism used for BKSF is similar to the stabilizer formalism used for error correction. 

The organization of the paper is as follows: in the next section, we review the background for quantum simulation.  This includes the algebra of fermionic operators, the electronic Hamiltonian in second quantization, and the phase estimation algorithm (PEA). 
Spin-to-fermion transforms are described in Section \ref{sec:Transformations} with an emphasis on BKSF.
The BKSF-transformed qubit operators corresponding to fermionic operators in the Hamiltonian are given in Section \ref{sec:Transformations}. In Section \ref{sec:hyd_mol}, we will illustrate the BKSF algorithm using a hydrogen molecule. The results and comparison of BKSF to other algorithms for the case of hydrogen molecule are presented in Section \ref{sec:Results}.

\section{Background for quantum simulation} \label{sec:TheoryBackground}

To explain quantum simulation, we first need to introduce fermionic systems and their Hamiltonians.  Throughout the paper we will be using second quantization where the algebra of the operators ensures antisymmetry of the electronic wave function.  Our work focuses on non-relativistic quantum simulation; however, note that relativistic quantum simulation has been addressed in \cite{Veis2011}. 

Section \ref{ss:fer_sys} will describe electronic Hamiltonian in second quantization and how to transform fermionic operators to qubit operators. The Trotter-Suzuki formulas for approximating unitary evolution are presented in Section \ref{ss:Trotter}. The phase estimation algorithm, described in Section \ref{ss: PEA}, can be used to extract the energy.  There are a number of related methods and approaches \cite{Childs2015,Mcclean2016,Paesani2017} that are known to have better asymptotic behavior.  However, given that this paper focuses on the BKSF algorithm, we will not delve into more sophisticated methods for time evolution and quantum measurement.

\subsection{Fermionic Systems and Second Quantization} \label{ss:fer_sys}
The evolution of a non-relativistic quantum system is governed by Schr\"odinger's equation, which in atomic units ($m_e=e=\hbar=1$), is given by
\begin{eqnarray}
i\frac{\partial \Psi}{\partial t} (\vec{\mathbf{x}},\vec{\omega},t)
&=&
H(\vec{\mathbf{x}})
\Psi(\vec{\mathbf{x}},\vec{\omega},t)
\end{eqnarray}
with
\begin{align}
H(\vec{\mathbf{x}})=
\left(- \sum_i^N \left[\frac{\nabla_{x_i}^2}{2} +V_{ext}(x_i) +\frac12\sum_{j}^{N} \frac{1}{\abs{x_i-x_j}}\right]
\right)
\label{elecHam}
\end{align}
Here $\vec{\mathbf{x}}=(x_1,x_2,...x_N)$ represents the spatial coordinates of all $N$ electrons and $\vec{\omega}=(\omega_1,...\omega_N)$ represents their spin coordinates with $\omega_j=\pm \frac12$.  Note that we assumed that the external electronic potential is constant in time which is consistent with the  approximation of static nuclei.

Our primary focus in this article is on electronic structure. Consequently, we will require that the $N$-body wave function $\Psi$ be completely antisymmetric.  To enforce antisymmetry, we will impose the fermionic algebra on the creation and annihilation operators, $\{a_j^\dag\}^M_{j=1}$ and $\{a_j\}^M_{j=1}$ respectively. 
In the second quantization formalism, the operators create and annihilate into single particle modes $\{\chi_j(x,\omega)\}_{j=1}^M$. These single particle wave functions are called modes or orbitals throughout this paper.

The antisymmetric property of the wave function, is built into the algebra by having the creation and annihilation operators satisfy following relations:
\begin{align}
a_ja_k+a_ka_j &=0,
&a_ja_k^{\dag}+a_k^{\dag} a_j&=\delta_{jk}\mathbf{1}
\end{align}
The action of creation and annihilation operators on occupation number $N$-body basis states $\ket{f_1...f_M}=(a_i^\dag)^{f_1}...(a_M^\dag)^{f_M}\ket{\Omega}$ is given by:
\begin{align}
a_j^\dag \ket{f_1...f_{j-1} 0 f_{j+1}...f_{M}}&= (-1)^{\Gamma_j}\ket{f_1...f_{j-1} 1 f_{j+1}...f_{M}}\nonumber\\
a_j^\dag \ket{f_{1}...f_{j-1} 1 f_{j+1}...f_M} &= 0\nonumber\\
a_j \ket{f_{1}...f_{j-1} 1 f_{j+1}...f_M} &= (-1)^{\Gamma_j}\ket{f_{1}...f_{j-1} 0 f_{j+1}...f_M}\nonumber\\
a_j \ket{f_{1}...f_{j-1} 0 f_{j+1}...f_M} &=0 \label{ac_relations}
\end{align}
Here $\Gamma_j=\sum_{s=1}^{j-1}f_s$, and $f_{i}$ represents the occupation of fermionic mode $\chi_{i}(x,\omega)$ and $\ket{\Omega}$ is the vacuum state with no fermionic particles present.

With these operators in hand, we can turn to expressing both the Hamiltonian and $N$-body wave function in terms of the underlying one-particle basis set. Firstly, we can write the Hamiltonian as
\begin{align}
H=\sum_{ij}^M h_{ij}a_{i}^{\dagger}a_{j}+\frac12 \sum_{ijkl}^M h_{ijkl}a_{i}^{\dagger}a_{j}^{\dagger}a_{k}a_{l}\label{hyd_ham}
\end{align}
with $h_{ij}$ and $h_{ijkl}$ one-electron and two-electron integrals of the operators projected into the basis set.  That is to say,
\begin{align}
h_{ij} &=\int\chi_{i}^{\ast}(x,\omega)\left(- \frac{\nabla^2_x}{2} +V_{ext}(x)\right)\chi_{j}(x,\omega)dx d\omega \nonumber \\
& =\delta_{\sigma_{i}\sigma_{j}}\int\phi_{i}^{\ast}(x) \left(- \frac{\nabla^2_x}{2} +V_{ext}(x)\right)\phi_{j}(x)dx \end{align}
and
\begin{align}
h_{ijkl} 
&=\delta_{\sigma_{i}\sigma_{l}}\delta_{\sigma_{j}\sigma_{k}} \int\frac{\phi_{i}^{\ast}(x_{1})\phi_{j}^{\ast}(x_{2})\phi_{k}(x_{2})\phi_{l}(x_{1})}{\abs{x_{1}-x_{2}}} dx_{1}dx_{2} 
\label{int_eq} 
\end{align}
Here, the modes within the basis set were decomposed as $\chi_i(x,\omega)=\phi_i(x)\sigma_i(\omega)$ with $\phi_i(x)$ as the spatial orbital and $\sigma_i(\omega)$ as the spin function. Since the Hamiltonian does not interact with the spin component of the wave function, $\sigma_i$ is either $\alpha(\omega)$ or $\beta(\omega)$ corresponding to spin up or spin down.  The two orthogonal spin functions are defined via $\alpha(\frac12)=\beta(-\frac12)=1$ and $\alpha(-\frac12)=\beta(\frac12)=0$.

The fermionic operators used in the second quantized representation of electronic Hamiltonian 
obey the fermionic algebra 
which is different from the algebra obeyed by qubits. Therefore, a transformation from fermionic operators to qubit operators is required to represent the Hamiltonian in terms of qubit operators. 

\subsection{Implementing qubit Hamiltonians on quantum computers}\label{ss:Trotter}

Qubit representation of the Hamiltonian can be used to time evolve a given state on a quantum computer. To implement the evolution, we need to implement the unitary, $\exp(-iHt)$. If all the qubit operators in the Hamiltonian commute then the time evolution of a given state under a given electronic Hamiltonian is relatively simple to achieve.  The exponential can be implemented as follows:
	\begin{align}
	\exp(-iHt)=\exp(-i\sum_{i} H_i t)=\prod\limits_{i}\exp(-iH_i t)
	\end{align}
	where $H$ is the Hamiltonian, and $H_{i}$ represents individual terms of $H$ with qubit operators. So, each of the qubit operator terms can be exponentiated in succession to implement the Hamiltonian.
    
    But, in general, the terms do not commute. Even in the case of hydrogen molecule in a minimal basis set, there are many non-commuting terms in the Hamiltonian. So, an approximation to the time evolution operator is required. The Trotter-Suzuki formula \cite{Trotter1959,Suzuki1992} is used to approximate the time evolution of the Hamiltonian. For two non-commuting operators $A$ and $S$, the first order Trotter-Suzuki formula is given as:
	\begin{align}
	e^{(A+S)t}&\approx\left(e^{\frac{At}{n}}e^{\frac{St}{n}}\right)^{n}+O(t(\delta t))
	\end{align}
    where, $t$ is the time for which time evolution takes place, $\delta t=t/n$ is the time step,  and $n$ is the number of Trotter steps used for the approximation. Increasing the number of Trotter steps decreases the error and hence, better approximates the original time evolution operator. 
	
    Another way to reduce the error is to use higher order Trotter-Suzuki formulas \cite{Trotter1959,Suzuki1992}. The second, third and fourth order Trotter-Suzuki formulas are given as:
    	\begin{align}
	e^{(A+S)t}\approx&\left(e^{\frac{At}{2n}}e^{\frac{St}{n}}e^{\frac{At}{2n}}\right)^{n}+O(t(\delta t)^{2})\\
	e^{(A+S)t}\approx&\left(e^{\frac{7At}{24n}}e^{\frac{2St}{3n}}e^{\frac{3At}{4n}}e^{\frac{-2St}{3n}}e^{\frac{-1At}{24n}}e^{\frac{St}{n}}\right)^{n}+O(t(\delta t)^{3})\\       	
	e^{(A+S)t}\approx&\left(\prod_{i=1}^{5}e^{\frac{p_{i}At}{2n}}e^{\frac{p_{i}St}{n}}e^{\frac{p_{i}At}{2n}}\right)^{n}+O(t(\delta t)^{4})
	\end{align}
	where
	$$p_1=p_2=p_4=p_5=\frac{1}{4-4^{\frac{1}{3}}}, \qquad p_3=1-4p_1$$	
    for the fourth order. In our study we focused on comparing the results of different algorithms using the first order Trotter-Suzuki formula. We present the results in the Section \ref{sec:Results}.
    
An alternative is the Taylor approximation method where
\begin{align}
\tilde{U}_{approx}=\sum^{x}_{n=0}(-it)^n H^{n}
\end{align}
is projected onto a unitary operator using amplitude amplification \cite{Childs2015}.

\subsection{Phase Estimation Algorithm} \label{ss: PEA} Given an unitary operator, $U$, and its eigenvector, $v$, the phase estimation algorithm \cite{Neilsen1998} allows us to calculate the eigenvalue, $\exp(2\pi i \phi)$ corresponding to the given eigenvector. To carry out the algorithm we need two registers, one containing the eigenvector, and another to store the eigenvalue. The schematic of the algorithm is as follows:
		\[
		\Qcircuit @C=.5em @R=0.4em @!R {
			\lstick{\ket{0}}& \qw {/}& \gate{QFT} & \ctrl{1} & \gate{QFT^{-1}} & \qw & \meter \\
			\lstick{\ket{v}}&\qw  {/}&\qw & \gate{U} & \qw &\qw\\
			}
			\] 
	The eigenstate, $\ket{v}$, is initialized and then $U$ is applied for a given amount of time. The eigenstate picks up a phase proportional to the time for which the unitary was applied and this phase is read from the first register. For the case of the electronic Hamiltonian, the unitary is $\exp(-iHt)$, and the eigenvector is the ground state. As discussed earlier, applying $\exp(-iHt)$ is not trivial for the Hamiltonian that contains non-commuting terms. Hence, the Trotter-Suzuki approximation presented in the last subsection, is used. The number of Trotter steps can be increased until chemical precision of $10^{-4}$ Hartrees is achieved.
    
\subsection{State preparation}

Although the phase estimation algorithm can be used to interrogate the state efficiently on quantum computers, it is not always obvious how to prepare the initial state of the simulation.  In fact, such state preparation problems can be formulated as representatives of the quantum NP class \cite{Kitaev2002}.  Despite the difficulty of the worst case problem instances, many practical instances can be approached using classical and quantum heuristics for state preparation. 

Classical approximations for the wave function with polynomial length fixed-basis expansions can be input to quantum computers \cite{Somma2002,Ortiz2001}.  Thus, Hartree-Fock wave functions \cite{Aspuru-Guzik2005,Whitfield2011} as well as post-Hartree-Fock wave functions \cite{Veis2010,Wang2008} can be used as approximate ground states.  If the approximate state has polynomial overlap with the ground state, the phase estimation algorithm can then be used a polynomial number of times to project the approximate state into the ground state.

In the cases that the classical approximation is insufficient, quantum heuristics have also been developed to assist with state preparation.  Adiabatic state preparation \cite{Aspuru-Guzik2005,Lidar1999} starts with a Hamiltonian with an easy to prepare ground state and slowly changes it to the electronic Hamiltonian of interest.  The rate of change should be slow to satisfy the adiabatic theorem, however it is usually not known beforehand what the optimal schedule is.  

Another quantum state preparation and measurement technique is the variational quantum eigensolvers (VQE) \cite{Peruzzo2014,Mcclean2016,Romero2017}.  The VQE allows a feedback loop between the classical and quantum computer to be established \cite{Mcclean2016} and has been used in experimental investigations \cite{Omalley2016}.

\subsection{OpenFermion} \label{ss:OF} To facilitate the wide-spread adoption of quantum simulation, researchers have been developing a Python-based toolset called OpenFermion.  OpenFermion (previously known as Fermilib) \cite{Mcclean2017} is an open source project providing the classical routines necessary for simulating fermionic models and quantum chemistry problems on quantum computers. It is written mostly in Python and supports ProjectQ \cite{Steiger2016}, Psi4 \cite{Psi42017} and few other open source projects for quantum chemistry problems. ProjectQ provides an interface for hardware and Psi4 provides an interface for standard electronic structure routines such as Gaussian integral evaluations. 

The quantum simulation of a molecule begins with Hamiltonian integrals computed using Psi4. These Hamiltonian parameters are transformed to the qubit operators, e.g. via the BKSF algorithm, using OpenFermion. The qubit Hamiltonian can be further compiled to quantum hardware such as e.g. for the IBM Quantum Experience~\cite{Cross2017}.  The hardware interfacing can be done directly through OpenFermion or via ProjectQ. 

To validate the algebraic expressions of different operators in BKSF, Matlab code was developed for both BKSF and Jordan-Wigner. After validation, the code was ported to Python and contributed to the OpenFermion project. Most of the work presented in this paper including full implementation of BKSF and extensive test routines can be found at the OpenFermion repository \cite{Mcclean2017}. 

\section{Transformations from fermionic operators to qubit operators} \label{sec:Transformations}
One of the main steps of quantum simulation and the central idea for this paper is the mapping from fermionic operators to qubits operators. Given the coherence times of current quantum computers, it is crucial to have an efficient mapping. The mapping affects efficiency with which time evolution under a given Hamiltonian can be implemented. This section will describe Jordan-Wigner and Bravyi-Kitaev transformation briefly. BKSF will be explained in detail.

\subsection{Jordan-Wigner and Bravyi-Kitaev Transformation}
Just as classical computers use two-level bits as their basic units, quantum computers use qubits as their basic units which are also two-level systems. There exists an useful correspondence between orbitals and qubits, since the occupation number of a particular orbital can be associated with the two states of the qubits. Direct identification between occupation of orbitals and the two levels of qubits, was first defined in the context of 1D lattice models \cite{Jordan1928} and then proposed as a scheme for simulating fermions \cite{Ortiz2001}. 

The raising and lowering operators for qubits, given as:
\begin{align}
Q^{\pm}=\frac{1}{2}(\sigma^{x}\mp i\sigma^{y})
\end{align}
do not satisfy the anti-commutation relations described by Eq. \eqref{ac_relations}. This motivates the following definition for creation and annihilation operators:
\begin{align}
a_j^{\dag}\equiv\mathbf{1}^{\otimes n-j-1}\otimes Q_{j}^{+}\otimes[\sigma^z]^{\otimes j}\\
a_j\equiv\mathbf{1}^{\otimes n-j-1}\otimes Q_{j}^{-}\otimes[\sigma^z]^{\otimes j}
\end{align}
With these operators, the fermionic Hamiltonian represented in second quantization can be transformed to the qubit Hamiltonian. 

Analyzing the expressions above, it is seen that the locality of a single fermionic operation scales as $O(M)$ where $M$ is the number of modes. This is because for an operation with the creation or annihilation operator, the qubits preceding it are also involved to get the correct parity (sign). This $O(M)$ cost can turn out to be quite expensive as $M$ gets large. The circuit for the transformed Hamiltonian can be processed before actually implementing it on a quantum computer. This has been explored by \cite{Hastings2015} and the cost can be brought down.

Another way to improve the simulation is by using a mapping that lowers the cost of creation and annihilation operators. If instead of occupation numbers, the parity is stored in the qubits then the cost for the parity operator would come down from $O(M)$ to $O(1)$. But, with this parity scheme, every time a fermionic operator is applied it will be required to update all the qubits following it to store the right parity. So, there is no improvement and the cost for the fermionic operators remain $O(M)$. The alternative to this was provided by Bravyi and Kitaev \cite{Bravyi2000} and had already been explored from computer science perspective \cite{Fenwick1995}. The idea is to strike a balance between storing parity and occupation number in the qubits. It is similar to idea of Fenwick trees \cite{Fenwick1995,Havlicek2017} in classical computation. With Fenwick trees, occupation number basis are mapped to different basis such that the cost of parity operators and update operators is $\log(n)$.
In essence, Bravyi-Kitaev is a compromise between the Jordan-Wigner and parity-scheme.

\subsection{Bravyi-Kitaev Superfast (BKSF) Algorithm}
BKSF defines another spin-fermion mapping where the number of gates required to implement fermion operators are $O(d)$, where $d$ is a constant.
This model draws some inspiration from Kitaev's honeycomb model \cite{Kitaev2005, Pachos2007} where Majorana modes are used to make the diagonalization of the Hamiltonian tractable. First proposed by Ettore Majorana \cite{Majorana2006}, the Majorana modes were real solutions to the Dirac equation. Each modes can, at least mathematically, be split into two Majorana modes. Majorana modes can be defined as follows:
\begin{align}
c_{2j}&= a_{j}+a^{\dagger}_{j} \\
c_{2j+1}&= -\ii(a_{j}-a^{\dagger}_{j})
\end{align}
These operators are Hermitian and satisfy:
\begin{align}
c_{j}c_{k}+c_{k}c_{j}=2\delta_{jk} \label{eq:Maj_modes}
\end{align}
With the above definition, Majorana fermions should exist in all fermionic systems. 

	An interesting approach, first presented by Kitaev in the context of quantum memory, was to pair Majorana fermions by interaction \cite{Kitaev2001}. This would lead to few unpaired Majorana modes which cannot interact with environment on their own because the operator, $c_{j}$ is not physical. And hence information encoded in such a system would stay protected from the environment. 

The BKSF model uses Majorana modes as intermediate operators corresponding to the interaction of fermionic modes, which are then transformed to qubit operators. Unlike the Jordan-Wigner and Bravyi-Kitaev algorithms, the BKSF algorithm requires the mapping of problem to a new model where more qubits are required. 

A brief summary of the BKSF algorithms is:
	\begin{itemize}
		\item[1)] Map the orbitals/modes to vertices of a graph. Number of modes required are equal to the number of one particle basis set functions used to expand Hamiltonian. 
		\item[2)] Given the Hamiltonian, get the number of edges required for the model.
		\item[3)] Put the qubits on the edges and find the edge operator expressions for terms in the Hamiltonian using expressions presented in Section \ref{sec:BKSF_rep}.
		\item[4)] Find the qubit operator representation of edge operators using Eq. \eqref{B_i} and \eqref{A_ij}. In the Hamiltonian, replace the edge operators, $A_{ij}$ and $B_{i}$ with $\tilde{A}_{ij}$ and $\tilde{B}_{i}$. 
		\item[5)] Find the independent loops in the graph. Define stabilizers for those loops.
		\item[6)] Use the stabilizers to find the relevant subspace for quantum simulation and the vacuum state.
		\end{itemize}
		After these steps the Hamiltonian expressed in qubit operators can be used along with the phase estimation algorithm (PEA) to find the energy of the molecule. We will now present details about the BKSF algorithm.

BKSF algorithm defines an abstract model by mapping fermionic modes to the vertices of a graph. If two modes interact then an edge is put between the vertices corresponding to the modes. If there is no interaction term in the Hamiltonian for the two modes, then the corresponding edge in not required in the graph. After constructing the graph corresponding to the Hamiltonian, qubits are put on the edges and new edge operators are defined which can generate the algebra of modes on the vertices. These operators are defined in terms of Majorana modes. 

For a case with $M$ modes, the graph will have $M$ vertices and let us assume that we have $E$, edges. Now, corresponding to the creation and annihilation operators, $a^{\dagger}_j$ and $a_{j}$, for the vertices, edge operators are defined as follows \cite{Bravyi2000}:
\begin{align}
	B_i&=-\ii c_{2i}c_{2i+1} \quad \text{for each vertex,}\\
	A_{ij}&=-\ii c_{2i}c_{2j} \quad \text{for each edge }(i,j) \in E,	
\end{align}
where $c_{2j}$ and $c_{2j+1}$ are the Majorana modes \eqref{eq:Maj_modes}. The edge operators satisfy the following relations \cite{Bravyi2000}:	
\begin{align*}
 & B^{\dagger}_{i}=B_{i},\qquad \qquad \quad \quad  \qquad A^{\dagger}_{ij}=A_{ij}, \\
& B_{i}^{2}=1,\qquad \qquad \quad \qquad \qquad A_{ij}^{2}=1, \\
 &B_{i}B_{j}=B_{j}B_{i}, \qquad \quad \qquad \quad  A_{ij}=-A_{ji} \\
 &\quad \quad A_{ij}B_{k}=(-1)^{\delta_{ik}+\delta_{jk}}B_{k}A_{ij}, \\
 &\quad A_{ij}A_{kl}=(-1)^{\delta_{ik}+\delta_{il}+\delta_{jk}+\delta_{jl}}A_{kl}A_{ij} \\
 &\quad \ii^{p}A_{j_{0}j_{1}}A_{j_{1}j_{2}}....A_{j_{p-2}j_{p-1}}A_{j_{p-1}j_{0}}=1 \numberthis \label{eq:stab_relation}\\
 &\text{for any closed loop path on the graph} \nonumber
\end{align*}

As will be shown in Section \ref{sec:BKSF_rep}, all the physical fermionic operators can be represented in terms of these edge operators. In essence, the edge operators can be used to generate fermionic algebra.

Furthermore, in this model, only the sector of the Hamiltonian that has an even number of particles is examined. This is because, with the two edge operators available to us, $A_{ij}$ and $B_{i}$, there is no combination of $A_{ij}$ and $B_{i}$ that produces single/odd number of Majorana mode operators. Similarly, an individual or an odd number of creation and annihilation operator represented using Majorana modes involve odd number of Majorana modes. So, we can establish that we cannot represent single/odd number of creation and annihilation operators in the BKSF model. But this is not a problem for physical operators because all the physical operators contain even multiples of creation and annihilation operators which lead to an even number of Majorana mode operators. So, the edge operators are sufficient to represent the electronic Hamiltonian and the model can be used to look at the even sector. For the even sector of the Hamiltonian the model satisfies another constraint:
\begin{equation}
\prod_{i} B_i=1
\end{equation} 	

We can sketch a way to simulate odd number of particles, but we have not included the details in our discussion. The Hilbert space of the abstract model defined for the Hamiltonian can be expanded by adding an extra mode. We can make sure that the added mode does not interact with the rest of the modes by not having any terms in the Hamiltonian that include the energy for this mode. The even sector of this new model can be used to simulate the odd sector of the original model. 

All the relations presented for the edge operators are in terms of Majorana modes. These Majorana modes can be mapped to qubits on the edges of the graph defined for the Hamiltonian. Then, the expressions of these edge operators in terms of the qubit operators can be calculated. The number of edge operators required for the complete description of the system would depend on the interaction which also determines the number of edge qubits required. In the smallest connected graph, which would be a closed chain, the number of edges are one less than the number of vertices. So, the Hilbert space of the qubits would actually be smaller than the Hilbert space of the qubits used in Jordan-Wigner or Bravyi-Kitaev transformation. But, as we are just considering the even particle sector, the Hilbert space associated with edge qubits is equal to the Fock space of even number of fermions.

In an open chain, the number of edges will be the same as the vertices. In general case, graphs will have more number of edges than vertices. So, the size of the Hilbert space spanned by the qubits on the edges will in general be bigger than the size of the Fock space of modes and the Hilbert space of qubits on the vertices.

To simulate the correct dynamics, one can define a code-space, a subspace of the Hilbert space of edge qubits, where the algebra of the edge operators holds. The construction of the subspace will be presented in the next section. But first, it is necessary to construct the expressions for edge operators in terms of qubit operators that satisfy the algebra. So, we associate the edge operators defined above with $\tilde{A}_{ij}$ and $\tilde{B}_{i}$ \cite{Bravyi2000}.

To define the new operators, we need to define the orientation of the graph. So we define an edge matrix $G$ where the entries are $\epsilon_{ij}=\pm 1$ if the edge is present and $0$ if the edge is not present. Further, $\epsilon_{ij}=-\epsilon_{ji}$ and we choose to $\epsilon=+1$ for $i<j$ and assume the edges for a particular vertex are ordered too. With this convention the edge operators are defined as:
\begin{equation}
\tilde B_i=\prod_{j: (ij)\in E}\sigma^{z}_{ij} \label{B_i}
\end{equation} and 
\begin{equation}
\tilde A_{ij}=\epsilon_{ij} \sigma^{x}_{ij}\prod_{l<j}^{n(i)} \sigma^{z}_{li} \prod_{s<i}^{n(j)} \sigma^{z}_{sj}\label{A_ij}
\end{equation}
where, $n(i)$ and $n(j)$ are the number of edges connected to vertices $i$ and $j$, respectively.

It can be checked that these operators satisfy all the relations satisfied by the original edge operators except for the product of $A_{ij}$ giving $\mathbf{1}$ on a closed loops. But, as we mentioned earlier the Hilbert space spanned by edge qubits is much bigger than the Hilbert space of the fermionic modes we started with. Therefore, if we can restrict ourselves to the subspace where the original relations hold then the mapping will be complete. 

The way to restrict to the subspace is through the use of stabilizers. The name stabilizer comes from the error correction where the stabilizer operators are used in similar sense to restrict states to the code-space \cite{Gottesman1997}. Here, we use stabilizers to remain in the Fock space of the fermions being simulated. Whether we can use the same construct to also achieve error correction while simulating fermions is something we would like to explore in future.			

\subsection{Stabilizers and Vacuum State}\label{ss:stab} As mentioned in the previous section, the edge operators $\tilde{A}_{ij}$ and $\tilde{B}_{i}$ defined, satisfy all the relations satisfied by $A_{ij}$ and $B_{i}$ except for Eq. \eqref{eq:stab_relation}. So, we define stabilizer as an operator $C_{L}$ such that:
\begin{align}
C_{L}= \ii^{n}A_{j_{1}j_{2}}A_{j_{2}j_{3}}....A_{j_{n-1}j_{n}}&A_{j_{n}j_{1}}, 
 \\
& L\in \left\lbrace \text{Loops in} (M,E) \right\rbrace		\nonumber
\end{align}
where $L$ represents any closed loop in the graph. And then we define the codespace, $\varepsilon$ as: 
\begin{align}
\varepsilon \equiv \left\{ \ket{\Psi}\enspace | \enspace C_{i}\ket{\Psi}= \ket{\Psi}\right\}
\end{align}			
If we look at the set of all the stabilizers then it can be check that stabilizer operators commute, which means it is possible to simultaneously diagonalize the set of stabilizer operators. So, the important question becomes, do we need to stabilize for all the operators, which in turn is equal to the total number of loops? And the answer turns out to be no. We just need to stabilize for all the linearly independent loops because as we prove next all the bigger loops are automatically stabilized if all the sub loops are stabilized. 			
\begin{figure}
	\includegraphics[width=8cm]{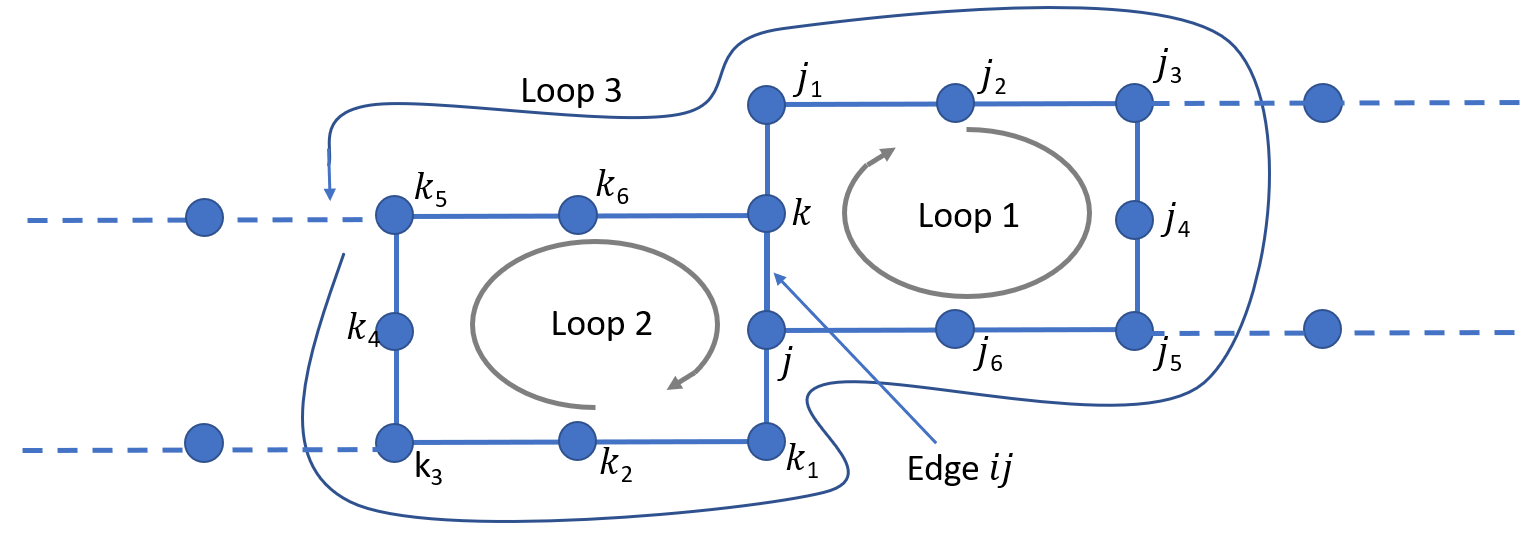}
	\caption{General graph showing a loop made up of two smaller loops. As we prove in the text, it is enough to stabilize the smaller independent loops and the bigger loop is automatically stabilized.}
    \label{stab}
\end{figure}

We consider a case where we have two small loops, e.g. $L_1$ and $L_2$ in the Figure \ref{stab}, and they share a common edge, which means there exists a bigger loop, $L_3$. What we are interested in is does stabilizing $L_1$ and $L_2$, stabilizes $L_3$? Let us look at the expression for the stabilizer for loop $L_1$ and $L_2$.
\begin{align*}
C_{L_1}= \ii^{8} A_{k j_{1}} A_{j_1j_2} ... A_{j_6 j} A_{jk} \\
C_{L_2}= \ii^{8} A_{kj} A_{jk_1}A_{k_1 k_2}  ... A_{k_6 k} 
\end{align*}			

One important choice that we make here is to end the stabilizer for  $L_1$ on the edge $jk$ as shown in the figure. Similarly we choose to start at edge $kj$ for  $L_2$. Here, in the expressions given above $k_{1}, k_2, j_1, j_2$ are indices for the vertices (see figure \ref{stab}). A few important points to emphasize before moving forward:
\begin{itemize}
	\item[1)]
    In the loops shown in the figure the smaller independent loops are both clockwise but we could just as well have chosen both anti-clockwise direction. We just need to make sure that the independent loops are all either clockwise or all anti-clockwise. It can be checked that, given the edge operator algebra the two directions are equivalent.
	\item[2)]The starting point of the loop also does not matter. We can choose to start at any edge. So, starting with edge $kj$ in loop $L_1$ does not result in the loss of generality.
\end{itemize}
Now, we flip the direction of $L_2$, and still make sure that in the expression for stabilizer for $L_2$ we end with $A_{kj}$ So, if look at the expression for $C_{L_1} C_{L_2}$ then we get
\begin{align}
C_{L_1}C_{L_2}&= \left(\ii^{8}A_{k j_{1}} A_{j_1j_2} ... A_{jk}\right)\left( \ii^{8}A_{kj}A_{jk_1}...A_{k_6 k} \right) \\
C_{L_1}C_{L_2}&= \ii^{16}A_{kj_{1}}A_{j_1j_2}... A_{j_6 j}\underbrace{A_{jk}A_{kj}}_{A^2_{jk}A_{kj}=-1}A_{jk_1}...A_{k_6 k}  \\
&= \ii^{14}A_{kj_{1}}... A_{j_6 j}A_{jk_1} A_{k_6k}=C_{L_3}
\end{align}
which means, given
\begin{align*}
C_{L_1}\ket{\Psi}=\ket{\Psi}, & \enspace C_{L_2}\ket{\Psi}=\ket{\Psi} \implies C_{L_3}\ket{\Psi}=\ket{\Psi}\numberthis
\end{align*}

The above proof can be generalized and applied to two or more loops with one or more common edges. The algebra of the edge operator ensures that we can construct the stabilizers for bigger loops from smaller sub loops. In showing that stabilizing smaller loops stabilizes the bigger loops we just made use of algebra of edge operators which would also hold in case of associated operators $\tilde{A}_{ij}$ and $\tilde{B}_i$ and hence is valid for the new edge operators too. The next important question is, for a given graph how many loops do we have to consider? The number of independent loops turn out to be $E-M+1$, and there have been many efficient classical algorithms to figure the independent loops from a given graph \cite{Paton1969}.

One important property of stabilizers is that they can also be used as projectors to the code space. So, if we start with a given state, not necessarily in the code-space, then we can construct an operator to project it to the code-space:
\begin{align}
\prod_{i=1}^{E-M+1} \frac{(1+C_{L_{i}})}{\mathcal{N}}\ket{\psi} \label{vacuum}
\end{align}
where the product is over all the independent loops, $\mathcal{N}$ is a normalization constant and $\psi$ is any state. If we start with the state with all the qubits in the $\ket{0}$ state and then use the stabilizers to project it to the code space then we wind up with vacuum state. This can be checked by acting with number operator on the vacuum state to see that it has zero particles. Vacuum state is very important from the point of view of quantum simulation since we need it to generate the initial number of fermions for simulation. The expressions for both the number operator and the operator with which we can initialize a state, in terms of the edge operators are given in the next section.

\section{BKSF representation of electronic Hamiltonian} \label{sec:BKSF_rep}
This section presents the necessary algebraic expressions for various operators in the electronic Hamiltonian required to perform the BKSF transformation. The algebraic expressions will be represented in terms of the edge operators. The Pauli-representation of the edge operators are used to get the Pauli-representation of  the Hamiltonian in BKSF basis.

The fermionic Hamiltonian has terms/operator that can be categorized into five types. The five types of operators and their algebraic expressions in terms of creation and annihilation operators are shown in Table \ref{Table:operators}. These operators need to be expressed in terms of the edge operators, following which the corresponding qubit edge operators $\tilde{A}_{ij}$ and $\tilde{B}_{i}$, can be used to get the Pauli representation of Hamiltonian. The properties of Majorana fermions are:
\begin{align*}
 c_{2i}=a_{i}+a^{\dagger}_{i}, \quad&\quad  c_{2i+1}=-\ii(a_{i}-a^{\dagger}_{i}), \qquad c_{2i}^2=1,\\
   &c_{i}c_{j}+c_{j}c_{i}=2\delta_{ij}
\end{align*}

Each class of fermionic operators are expressed in terms of edge operators next.\\
 
\begin{table}
	\caption{Second quantization form for different fermionic operators}
	\label{Table:operators}
	\centering
	\begin{tabular}{c c}
		\hline
		\hline
		\textbf{Operator} & \textbf{Second quantized form} \\
		\hline
		Number operator & $h_{ii}a^{\dagger}_{i}a_{i}$\\
		Coulomb/exchange operators & $h_{ijji}a^{\dagger}_{i}a^{\dagger}_{j}a_{j}a_{i}$\\
		Excitation operator & $h_{ij}(a^{\dagger}_{i}a_{j}+a^{\dagger}_ja_{i})$\\
		Number-excitation operator & $h_{ijjk}(a^{\dagger}_{i}a^{\dagger}_{j}a_{j}a_{k}+a^{\dagger}_{k}a^{\dagger}_{j}a_{j}a_{i})$\\
		Double excitation operator & $h_{ijkl}(a^{\dagger}_{i}a^{\dagger}_{j}a_{k}a_{l}+a^{\dagger}_{l}a^{\dagger}_{k}a_{j}a_{i})$\\
		\hline
		\hline
	\end{tabular}
\end{table}

\textbf{Number operator } This operator shows up in the Hamiltonian in context of potential energy in the position basis. The occupation number operator provides the information about whether the mode is occupied or not. If the mode is occupied then potential energy of that particular modes is taken into account while calculating the total energy of the system. The fermionic number operator is given by $a_{i}^{\dagger}a_{i}$, where $i$ represent the modes number.  If we start with the edge operator, $B_{i}$, we get, 
\begin{align*}
B_{i}=&-\ii c_{2i}c_{2i+1}\\
=&-\ii (a_{i}+a^{\dagger}_{i})(-\ii (a_{i}-a^{\dagger}_{i}))\\						
=& a_{i}a^{\dagger}_{i}-a^{\dagger}_{i}a_{i}\\
=& 1-2a^{\dagger}_{i}a_{i} \\
\implies a^{\dagger}_{i}a_{i}=&\frac{1-B_{i}}{2}\numberthis
\end{align*}
	
	 \textbf{Coulomb/exchange operator} Coulomb operator and exchange operator have similar algebraic expressions but the integrals associated with the operators provide very different information. Coulomb operator checks if the two modes are occupied and if they are then associated Coulombic interaction energy is added to the total energy. It is difficult to associate a physical picture with the exchange operator which is consequence of antisymmetric nature of the fermionic wave function. The algebraic expression for Coulombic/exchange operator is $a^{\dagger}_{i}a^{\dagger}_{j}a_{j}a_{i}$ which can also be written as $a^{\dagger}_{i}a_{i}a^{\dagger}_{j}a_{j}$, since $a_{i}a_j^\dag a_{j}=a_j^\dag a_{j}a_{i}$. Therefore, we have
	\begin{align*}
	 a^{\dagger}_{i}a_{i}a^{\dagger}_{j}a_{j}&=(a^{\dagger}_{i}a_{i})(a^{\dagger}_{j}a_{j})\\
	& =\left(\frac{1-B_{i}}{2}\right)\left(\frac{1-B_{j}}{2}\right).\numberthis
	\end{align*}

	 \textbf{Excitation operator} Excitation operator provides information about the kinetic energy associated with fermions while going from one modes to the other. The expression for the excitation operator is given as $a_{i}^{\dagger}a_{j}+a_{j}^{\dagger}a_{i}$. We represent this expression in terms of Majorana modes as:
	\begin{align*}
	a^{\dagger}_{i}a_{j}&=\frac{(c_{2i}-\ii c_{2i+1})}{2}\frac{(c_{2j}+\ii c_{2j+1})}{2}\\
	&=\frac{1}{4}(c_{2i}c_{2j}+\ii c_{2i}c_{2j+1}-\ii c_{2i+1}c_{2j}+c_{2i+1}c_{2j+1})
	\end{align*}
	Similarly,
	\begin{align*}
	a^{\dagger}_{j}a_{i}&=\frac{(c_{2j}-\ii c_{2j+1})}{2}\frac{(c_{2i}+\ii c_{2i+1})}{2}\\
	&=\frac{1}{4}(c_{2j}c_{2i}+\ii c_{2j}c_{2i+1}-\ii c_{2j+1}c_{2i}+c_{2j+1}c_{2i+1})\\
	&=\frac{1}{4}(-c_{2i}c_{2j}+\ii c_{2j}c_{2i+1}+\ii c_{2i}c_{2j+1}-c_{2i+1}c_{2j+1})
	\end{align*}
	Adding the two expressions,
	\begin{align*}
	a^{\dagger}_{i}a_{j}+a^{\dagger}_{j}a_{i}&=\frac{1}{2}(\ii c_{2i}c_{2j+1}+\ii c_{2j}c_{2i+1})
	\end{align*}
	Therefore, we get
	\begin{align*}
	a^{\dagger}_{i}a_{j}+a^{\dagger}_{j}a_{i}&=\frac{1}{2}\left(\ii\frac{A_{ij}}{-\ii}\frac{B_{j}}{-\ii}+\ii\frac{A_{ji}}{-\ii}\frac{B_{i}}{-\ii}\right)\\
	&=\frac{-\ii}{2}(A_{ij}B_{j}+B_{i}A_{ij})\numberthis
	\end{align*}			
	 \textbf{Number excitation operator} Number excitation operator is a two-body term which arises because of the antisymmetric nature of the wave function. It is difficult to come up with physical interpretation of number excitation operator and double excitation operator which is presented next. The number excitation operator is given by:
	 $$a^{\dagger}_{i}a^{\dagger}_{j}a_{j}a_{k}+a^{\dagger}_{k}a^{\dagger}_{j}a_{j}a_{i}$$ 
	 which can be rearranged in terms of the number operator and excitation operator, for which we already have the expressions in terms of edge operators.
	\begin{align*}
	a^{\dagger}_{i}a^{\dagger}_{j}a_{j}a_{k}+a^{\dagger}_{k}a^{\dagger}_{j}a_{j}a_{i}&=(a^{\dagger}_{i}a_{k}+a^{\dagger}_{k}a_{i})(a^{\dagger}_{j}a_{j})\\
	&=\frac{-\ii}{2}(A_{ik}B_{k}+B_{i}A_{ik})\frac{(1-B_{j})}{2}\numberthis
	\end{align*}					
	 \textbf{Double excitation operator} For double excitation operator we take the usual approach where we put in the expressions for $a^{\dagger}_{i}$ and $a_{i}$ in terms of the Majorana modes and make use of the properties of Majorana modes to simplify the expression.
	 The double excitation operator is given by:
	 $$a^{\dagger}_{i}a^{\dagger}_{j}a_{k}a_{l}+a^{\dagger}_{l}a^{\dagger}_{k}a_{j}a_{i}$$ Let us start with $a^{\dagger}_{i}a^{\dagger}_{j}a_{k}a_{l}$:
	\begin{align*}
	&a^{\dagger}_{i}a^{\dagger}_{j}a_{k}a_{l}\\
	& =\frac{(c_{2i}-\ii c_{2i+1})}{2}\frac{(c_{2j}-\ii c_{2j+1})}{2}\frac{(c_{2k}+\ii c_{2k+1})}{2}\frac{(c_{2l}+\ii c_{2l+1})}{2}
	\end{align*}
	Similarly,
	\begin{align*}
	&a^{\dagger}_{l}a^{\dagger}_{k}a_{j}a_{i}\\
	&=\frac{(c_{2l}-\ii c_{2l+1})}{2}\frac{(c_{2k}-\ii c_{2k+1})}{2}\frac{(c_{2j}+\ii c_{2j+1})}{2}\frac{(c_{2i}+\ii c_{2i+1})}{2}
	\end{align*}
	Now, if we add the two expressions, we are left with:
	\begin{align*}
	& a^{\dagger}_{i}a^{\dagger}_{j}a_{k}a_{l}+a^{\dagger}_{l}a^{\dagger}_{k}a_{j}a_{i}\\
	&=\frac{1}{8}(-A_{ij}A_{kl}-A_{ij}A_{kl}B_kB_l+A_{ij}A_{kl}B_iB_k\\
	&\qquad\quad +A_{ij}A_{kl}B_{i}B_{l}+A_{ij}A_{kl}B_jB_k+A_{ij}A_{kl}B_jB_l \\
	&\qquad\quad-A_{ij}A_{kl}B_iB_j+A_{ij}A_{kl}B_iB_jB_kB_l)\\
	& =\frac{1}{8}A_{ij}A_{kl}(-1-B_iB_j+B_iB_k+B_iB_l+B_jB_k\\
	&\qquad\qquad\qquad+B_jB_l-B_kB_l+B_iB_jB_kB_l)\numberthis
	\end{align*}				
	\textbf{Particle creation operator} The way to initialize a state using fermionic creation operators is straightforward. Applying creation operator, corresponding to the occupied mode, to the vacuum initializes the state. But, as we have discussed the edge operators can not be used to represent single/odd number of creation or annihilation operators. So, initialization must always occur in pairs. 
    
    We can start with $a_{i}^{\dagger}a_{j}^{\dagger}$ and follow the same procedure where we substitute the creation and annihilation operators in terms of Majorana operators.And then we can try to simplify the expression to get it in terms of edge operators. Instead of this long procedure we can make a guess based on the expression for excitation operator:
	\begin{align}
	a_{i}^{\dagger}a_{j}^{\dagger}+a_{i}a_{j}=\frac{-\ii}{2}(A_{ij}B_{j}-B_{i}A_{ij})
	\end{align} 
	Adding annihilation operators to the expression is not a problem because acting on the vacuum with annihilation operators gives zero. And, it can be checked that the expression works.
	
	This completes all the analytic expressions we are going to need to represent any electronic Hamiltonian in BKSF basis. The edge operators, $A_{ij}$, $B_{i}$ used in all the above expressions can be replaced by the equivalent qubit operators $\tilde{A}_{ij}$ and $\tilde{B}_{i}$ for quantum simulation. Based on the derived expressions, only three terms lead to an edge between vertices in the graph:  excitation operator, number excitation operator and double excitation operator.
    
    Only when we have an $A_{ij}$ edge operator in an expression, we need to put an edge between vertices, $i$ and $j$. In the molecular electronic Hamiltonian the Coulomb interaction is pairwise and long distance.  However, the degree of the edge graph does not need to change when the terms are strictly Coulombic.  
    
	
\section{Electronic Hamiltonian of hydrogen molecule} \label{sec:hyd_mol}			
	We will use the abstract model presented in the previous section and apply it to the example of hydrogen molecule.  The electronic Hamiltonian, Eq. \ref{hyd_ham}, in minimal basis will have four fermionic modes and hence we will need four vertices for the graph. The Hamiltonian was worked out by \cite{Whitfield2011} and is given as follows:
	\begin{align*}
&H(1)=h_{11}a^{\dagger}_{1}a_{1}+h_{22}a^{\dagger}_{2}a_{2}+h_{33}a^{\dagger}_{3}a_{3}+h_{44}a^{\dagger}_{4}a_{4}\\	&H(2)=h_{1221}a^{\dagger}_{1}a^{\dagger}_{2}a_{2}a_{1}+h_{3443}a^{\dagger}_{3}a^{\dagger}_{4}a_{4}a_{3}+h_{1441}a^{\dagger}_{1}a^{\dagger}_{4}a_{4}a_{1}+\\
	& \qquad\qquad\qquad h_{2332}a^{\dagger}_{2}a^{\dagger}_{3}a_{3}a_{2}+(h_{1331}-h_{1313})a^{\dagger}_{1}a^{\dagger}_{3}a_{3}a_{1}\\
	&\qquad\qquad\qquad +(h_{2442}-h_{2424})a^{\dagger}_{2}a^{\dagger}_{4}a_{4}a_{2}\\
	&\qquad\qquad\qquad+h_{1243}(a^{\dagger}_{1}a^{\dagger}_{2}a_{4}a_{3}+a^{\dagger}_{3}a^{\dagger}_{4}a_{2}a_{1})\\
	& \qquad\qquad\qquad+h_{1423}(a^{\dagger}_{1}a^{\dagger}_{4}a_{2}a_{3}+a^{\dagger}_{3}a^{\dagger}_{2}a_{4}a_{1})\\
	&H=H(1)+H(2) \numberthis \label{eq:sec_quant_hyd_ham}
	\end{align*}
	This is the second quantization representation of the Hamiltonian in molecular orbital basis. The molecular orbitals are obtained using Hartree-Fock method \cite{Szabo1996}. Although, in this case the molecular orbitals can also be obtained based on the symmetry. The bonding and anti-bonding orbitals are even and odd combinations of the atomic orbitals. The one-electron and two-electron integrals can be computed using open-source quantum chemistry packages e.g. \cite{Psi42017,PySCF2017}. 
    
	Each of the fermionic terms in the Hamiltonian can be transformed to qubit operators using a particular transformation. The Jordan-Wigner and Bravyi-Kitaev transformation of hydrogen Hamiltonian has been detailed and analyzed in \cite{Seeley2012,Havlicek2017}. We will now present the steps required to transform the hydrogen molecule Hamiltonian from second quantization representation to the qubit operators in case of BKSF algorithm.  In BKSF, the orbitals are mapped to a connected and unoriented graph which can be constructed from the Hamiltonian.
	
\begin{figure}[h]
		\includegraphics[width=5cm]{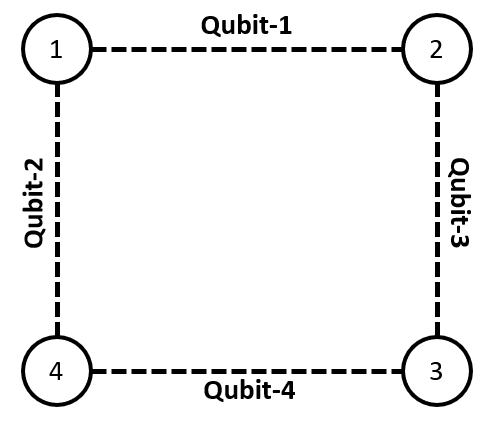}
		\caption{Pictorial representation of the fermionic modes and the interaction between them for the hydrogen molecule. This graph is based on the Hamiltonian given in Eq. \eqref{eq:hyd_H1} and \eqref{eq:hyd_H2}.}
        \label{fig:hydrogen_lfm}
	\end{figure}

	Based on the Hamiltonian, we will need four edges in the graph (note that the Coulomb terms do not contribute). The final graph is shown in the Figure \ref{fig:hydrogen_lfm}. Based on the graph, we would have four $\tilde{B}_i$ operators and four $\tilde{A}_{ij}$ operators.
	Before we give the Pauli representation of the edge qubits, the convention for the way qubits are labeled, needs to be specified. For the case of hydrogen where we have $4\cross4$ edge matrix the numbering would go something like, $$\overbrace{(3,4)}^{\circled{4}}\leftarrow\overbrace{(2,3)}^{\circled{3}}\leftarrow\overbrace{(1,4)}^{\circled{2}}\leftarrow\overbrace{(1,2)}^{\circled{1}}$$ 
	So, we would need four qubits in total to simulate. 
	
	The required operators for BKSF model in terms of Pauli operators are given below. We make use of Eq. \eqref{B_i} and \eqref{A_ij}, to get the Pauli representation of the edge operators.
	\begin{eqnarray}
	\label{A_B_tilde}
		& B_1=\sigma^{z}_2\sigma^{z}_1 \qquad\qquad\qquad& B_2=\sigma^{z}_3\sigma^{z}_1\nonumber \\
		& B_3=\sigma^{z}_4\sigma^{z}_3 \qquad\qquad\qquad& B_4=\sigma^{z}_4\sigma^{z}_2\nonumber \\		
		& A_{12}=\sigma^{x}_1	  \qquad\quad\qquad\qquad & A_{14}=\sigma^{x}_2\sigma^{z}_1 \nonumber \\
		& A_{23}=\sigma^{x}_3\sigma^{z}_1 \qquad\qquad\qquad &A_{34}=\sigma^{x}_4\sigma^{z}_3\sigma^z_2 \label{eq:A_B_tilde_end}
	\end{eqnarray}
	Next, we represent the hydrogen Hamiltonian in terms of BKSF edge operators. The one body terms in second quantization formulation are given by:
	\begin{align*}
H(1)&=h_{11}a^{\dagger}_{1}a_{1}+h_{22}a^{\dagger}_{2}a_{2}+h_{33}a^{\dagger}_{3}a_{3}+h_{44}a^{\dagger}_{4}a_{4}\numberthis \label{eq:hyd_H1}
	\end{align*}
	Substituting the number operator in terms of edge operators we get:
	\begin{align*}
	H(1)&=\frac{1}{2}(h_{11}(1-B_1)+h_{22}(1-B_2)+h_{33}(1-B_3)\\
	&\qquad+h_{44}(1-B_4)) \numberthis
	\end{align*}
	Similarly, two-body terms are given by:
	\begin{align*}
&H(2)=h_{1221}a^{\dagger}_{1}a^{\dagger}_{2}a_{2}a_{1}+h_{3443}a^{\dagger}_{3}a^{\dagger}_{4}a_{4}a_{3}+h_{1441}a^{\dagger}_{1}a^{\dagger}_{4}a_{4}a_{1}\\
	& \qquad\qquad\quad +h_{2332}a^{\dagger}_{2}a^{\dagger}_{3}a_{3}a_{2}+(h_{1331}-h_{1313})a^{\dagger}_{1}a^{\dagger}_{3}a_{3}a_{1}\\
	&\qquad\qquad\quad +(h_{2442}-h_{2424})a^{\dagger}_{2}a^{\dagger}_{4}a_{4}a_{2}\\
	&\qquad\qquad\quad+h_{1243}(a^{\dagger}_{1}a^{\dagger}_{2}a_{4}a_{3}+a^{\dagger}_{3}a^{\dagger}_{4}a_{2}a_{1})\\
	& \qquad\qquad\quad+h_{1423}(a^{\dagger}_{1}a^{\dagger}_{4}a_{2}a_{3}+a^{\dagger}_{3}a^{\dagger}_{2}a_{4}a_{1})\numberthis \label{eq:hyd_H2}       		
	\end{align*}
	substituting edge operators and simplifying we get:
	\begin{align*}
	H(2)=& \frac{h_{1221}}{4}(1-B_1-B_2+B_1B_2)\\
	&+\frac{h_{3443}}{4}(1-B_3-B_4+B_3B_4)\\
	&+\frac{h_{1441}}{4}(1-B_1-B_4+B_1B_4)\\
	&+\frac{h_{2332}}{4}(1-B_2-B_3+B_2B_3)\\
	&+\frac{(h_{1331}-h_{1313})}{4}(1-B_1-B_3+B_1B_3)\\
	&+\frac{(h_{2442}-h_{2424})}{4}(1-B_2-B_4+B_2B_4)\\
	&+\frac{h_{1243}}{8}A_{12}A_{43}(-1-B_1B_2+B_1B_4+B_1B_3\\
	&+B_2B_4+B_2B_3-B_4B_3+B_1B_2B_4B_3)\\
	&+\frac{h_{1423}}{8}A_{14}A_{23}(-1-B_1B_4+B_1B_2+B_1B_3\\
	&+B_4B_2+B_4B_3-B_2B_3+B_1B_4B_2B_3)			       		
	\end{align*}
	Now, replacing  $A_{ij}$ and $B_i$ with $\tilde{A}_{ij}$ and $\tilde{B}_i$, Eq. \eqref{eq:A_B_tilde_end} in $H(1)$ and $H(2)$ we get the Pauli representation of the Hamiltonian as:
	\begin{align*}
	H=& \frac{1}{4}(2h_{11}+2h_{22}+2h_{33}+2h_{44}+h_{1221}+h_{3443}\\
	&+h_{1441}+h_{2332}+h_{1331}-h_{1313}+h_{2442}-h_{2424})\\
	&+\frac{1}{4}(-2h_{11}-h_{1221}-h_{1441}-h_{1331}+h_{1313})\sigma^{z}_2\sigma^{z}_1 \\
	&+\frac{1}{4}(-2h_{22}-h_{1221}-h_{2332}-h_{2442}+h_{2424})\sigma^{z}_3\sigma^{z}_1\\
	&+\frac{1}{4}(-2h_{33}-h_{3443}-h_{2332}-h_{1331}+h_{1313})\sigma^{z}_4\sigma^{z}_3\\
    &+\frac{1}{4}(-2h_{44}-h_{3443}-h_{1441}-h_{2442}+h_{2424})\sigma^{z}_4\sigma^{z}_2\\
	&+\frac{1}{4}(h_{1221}+h_{3443})\sigma^{z}_3\sigma^{z}_2
	+\frac{1}{4}(h_{1441}+h_{2332})\sigma^{z}_4\sigma^{z}_1\\
	&+\frac{1}{4}(h_{1331}-h_{1313}+h_{2442}-h_{2424})\sigma^{z}_4\sigma^{z}_3\sigma^{z}_2\sigma^{z}_1\\
	&+\frac{h_{1243}}{4}(\sigma^{x}_4\sigma^{x}_1+\sigma^{y}_4\sigma^{y}_1+\sigma^{y}_4\sigma^{z}_3\sigma^{z}_2\sigma^{y}_1)\\       					&+\frac{h_{1423}}{4}(-\sigma^{z}_4\sigma^{x}_3\sigma^{x}_2\sigma^{z}_1-\sigma^{y}_3\sigma^{y}_2-\sigma^{z}_4\sigma^{y}_3\sigma^{y}_2\sigma^{z}_1)   \numberthis   \label{hyd_pauli} 
	\end{align*}
	This completes the transformation of the Hamiltonian to qubit operators. This Hamiltonian can be exponentiated and be used along with phase estimation algorithm (PEA) to get the energies of the molecule. 
    
    To restrict the dynamics to the Fock space, we need to make use of stabilizers. Stabilizers can be used to get the vacuum state of the system, which then can be used to initialize a given state. Once a state is initialized in the code space, it remains in code space as the edge operators commute with the stabilizer operators.
	
	As we discussed in the last section the stabilizers for a given Hamiltonian are based on the independent loops in the graph corresponding to the Hamiltonian. From Figure \ref{fig:hydrogen_lfm}, we can see that there is just one loop:
	\begin{align}
	L_1 \rightarrow (1234)
	\end{align}
	and it agrees with the formula for the number of independent loops, $L-M+1=1-1+1=1$. The expressions for the stabilizer for this loop is given by:
	\begin{align}
	Stab_{L_1}=A_{12}A_{23}A_{34}A_{41}
	\end{align}
	The edge operators $A_{ij}$ can be replaced by the corresponding $\tilde{A}_{ij}$ and Pauli representation for the stabilizers could be found. As we presented in Section \ref{ss:stab}, stabilizer operators define the code-space for the system. The stabilizer presented here will define the code-space for the hydrogen molecule system. We can find the vacuum state of the system in BKSF basis using Eq.~\eqref{vacuum}. Once the state is initialized using the vacuum state, the state will remain in the code space. 
    
\section{Results} \label{sec:Results}
In this section we start by presenting the Hamiltonian for hydrogen molecule in BKSF, Jordan-Wigner and Bravyi-Kitaev basis. Next, we present the number of gates required to implement a given term in the Hamiltonian. The number of gates required for a single Trotter step are presented in the Table \ref{Table:gatecount}. Finally, we present the analysis of Jordan-Wigner, Bravyi-Kitaev, and BKSF algorithm for the case of hydrogen molecule. 

To obtain the spin Hamiltonian, we used a combination of Psi4 and OpenFermion. The necessary molecular integrals are given in Table \ref{Table:integrals}.  First, the hydrogen molecular Hamiltonian via BKSF is given by:
 	\begin{align*}
	H_{BKSF}=& -0.045321\sigma^{z}_4\sigma^{x}_3\sigma^{x}_2\sigma^{z}_1-0.045321\sigma^{z}_4\sigma^{y}_3\sigma^{y}_2\sigma^{z}_1\\
&+0.171201\sigma^{z}_2\sigma^{z}_1+0.045321\sigma^{x}_4\sigma^{x}_1\\
&+0.171201\sigma^{z}_3\sigma^{z}_1 +0.3429725\sigma^{z}_3\sigma^{z}_2\\
&+0.045321\sigma^{y}_4\sigma^{y}_1 -  0.2227965\sigma^{z}_4\sigma^{z}_3\\ &+0.045321\sigma^{y}_4\sigma^{z}_3\sigma^{z}_2\sigma^{y}_1 -0.2227965\sigma^{z}_4\sigma^{z}_2\\
  &-0.045321\sigma^{y}_3\sigma^{y}_2   +0.2410925\sigma^{z}_4\sigma^{z}_3\sigma^{z}_2\sigma^{z}_1\\
    & +0.331736\sigma^{z}_4\sigma^{z}_1-0.812610I\numberthis \label{eq:H_BKSF}
	\end{align*}
\begin{table}
	\caption{One-electron and two-electron molecular integrals for hydrogen molecule. The integrals were calculated using the Psi4 with STO-3G \cite{Collins1976} as the basis set, and bond length 0.7414 $\AA$.}
	\label{Table:integrals}
	\centering
	\begin{tabular}{c c}
		\hline
		\hline
		\textbf{Integrals} & \textbf{Value (in Atomic Units)} \\
		\hline
		$h_{11}=h_{22}$ & -1.25246357\\[1ex]
		$h_{33}=h_{44}$ & -0.47594871\\[1ex]
		$h_{1221}=h_{2112}$ & 0.33724438\\[1ex]
		$h_{3443}=h_{4334}$ & 0.34869688\\[1ex]
		$h_{1331}=h_{1441}=h_{2332}=h_{2442}$ & 0.33173404\\
		$=h_{3113}=h_{4114}=h_{3223}=h_{4224}$ & \\[1.5ex]
		$h_{1313}=h_{2424}=h_{3241}=h_{3421}$ & 0.09064440\\
		$=h_{1423}=h_{1243}$&\\
		\hline
		\hline
	\end{tabular}
\end{table} 
The qubit Hamiltonians corresponding to the Jordan-Wigner and Bravyi-Kitaev algorithms are given by:
\begin{align*}
H_{JW}=&-0.81261I-0.045321\sigma^{x}_4\sigma^{x}_3\sigma^{y}_2\sigma^{y}_1\\
&-0.222796\sigma^{z}_3 +0.045321\sigma^{x}_4\sigma^{y}_3\sigma^{y}_2\sigma^{x}_1 \\	
&-0.222796\sigma^{z}_4+0.045321\sigma^{y}_4\sigma^{x}_3\sigma^{x}_2\sigma^{y}_1 \\
&+0.17434925\sigma^{z}_4\sigma^{z}_3-0.045321\sigma^{y}_4\sigma^{y}_3\sigma^{x}_2\sigma^{x}_1\\
& +0.171201\sigma^{z}_1+0.171201\sigma^{z}_2\\ 
&+0.1686232\sigma^{z}_2\sigma^{z}_1+0.165868\sigma^{z}_3\sigma^{z}_2\\
&+0.165868\sigma^{z}_4\sigma^{z}_1+0.120546\sigma^{z}_3\sigma^{z}_1\\
&+0.120546\sigma^{z}_4\sigma^{z}_2 \numberthis \label{eq:H_JW}
\end{align*} 
\begin{align*}
H_{BK}=&-0.81261I+0.045321\sigma^{x}_3\sigma^{z}_2\sigma^{x}_1 \\
 		 &-0.222796\sigma^{z}_3 +0.045321\sigma^{y}_3\sigma^{z}_2\sigma^{y}_1\\ 
	     &-0.222796\sigma^{z}_4\sigma^z_3\sigma^z_2+0.045321 \sigma^{z}_4\sigma^{x}_3\sigma^{z}_2\sigma^{x}_1\\ 
&+0.17434925\sigma^{z}_4\sigma^{z}_2+0.045321\sigma^{z}_4\sigma^{y}_3\sigma^{z}_2\sigma^{y}_1\\
&+0.171201\sigma^{z}_2\sigma^z_{1}+ 0.171201\sigma^{z}_1\\
&+0.1686232\sigma^{z}_2+0.165868\sigma^{z}_3\sigma^{z}_2\sigma^z_1\\
&+0.165868\sigma^{z}_4\sigma^{z}_3\sigma^z_2\sigma^z_1+0.120546\sigma^{z}_3\sigma^{z}_1\\
&+0.120546\sigma^{z}_4\sigma^{z}_3\sigma^z_1\numberthis \label{eq:H_BK}
\end{align*} 

To analyze total gate requirement for phase estimation algorithm, we need to consider the gate requirement for a given term in the Hamiltonian. So, let us consider the term $\sigma^z_3\otimes\sigma^z_2\otimes\sigma^z_1$. The circuit is given by:
\[
\Qcircuit @C=.5em @R=0em @!R {
	& \ctrl{1} & \qw & \qw & \qw & \ctrl{1}&\qw\\
	& \targ & \ctrl{1} & \qw & \ctrl{1} & \targ&\qw\\
	& \qw & \targ & \gate{R_{z}} & \targ & \qw&\qw
}
\]
The total number of gates required are five: four $CNOT$ gates, and one $R_z$. For $\exp(\ii X\otimes Y \otimes Z)$ we first change the basis and then apply $CNOT$ and $R_z$ gates and undo the change of basis.
\[
\Qcircuit @C=.5em @R=0.3em @!R {
	&\qw&\gate{H}& \ctrl{1} & \qw & \qw & \qw & \ctrl{1}&\gate{H}&\qw\\
	&\qw&\gate{R_{x}}& \targ & \ctrl{1} & \qw & \ctrl{1} & \targ&\gate{R_{x}^\dag}&\qw\\
	&\qw&\qw & \qw & \targ & \gate{R_{z}} & \targ & \qw&\qw&\qw
}
\]
Knowing the gate requirements for the two circuits presented, we can calculate the total gate requirement for each term in the Hamiltonian. Table \ref{Table:gatecount} presents the total gates required to implement the unitary, $\exp(-\ii Ht)$, using a single Trotter step for all the three algorithms. 
\begin{table}
	\caption{Total gate requirements to implement the unitary $\exp(-\ii Ht)$ for Jordan-Wigner, Bravyi-Kitaev, and BKSF algorithm.}
	\label{Table:gatecount}
	\centering
	\begin{tabular}{ p{6cm} p{1.7cm}}
		\hline
		\hline
		\textbf{Hamiltonian} & \textbf{Gate Count} \\
		\hline
		$H_{JW}$ & \qquad\quad82\\
		$H_{BK}$ & \qquad\quad74\\
		$H_{BKSF}$ & \qquad\quad79\\
		\hline
		\hline
	\end{tabular}
\end{table} 

Before continuing to describe our results, it should be noted that we have ignored the hardware implementation specific details. For example, it may not be possible to apply $CNOT$ gate between two qubits which are far from each other on an actual computer. In that case, we would have to use $SWAP$ gates adding to both the error and cost of the algorithm.  Alternatively, one could achieve non-local interactions using a mediating quantum bus qubit e.g.~\cite{Zhu17}.  Analyzing all the available architectures and coming up with optimum gate count for all of them is difficult task. So, we will just analyze the minimum number of gates that will be required, no matter the quantum computer architecture. 

Since in general, there are non-commuting terms in the Hamiltonian, implementation of the unitary $\exp(-iHt)$ is not as straightforward. We need to use Trotter-Suzuki approximation as presented in Section \ref{ss:Trotter}. In this initial investigation into the BKSF, we have used first order Trotter-Suzuki approximation for our analysis. Trotter-Suzuki approximation depends on the order in which the terms of the Hamiltonian are implemented. If the ordering is not optimized then it can take up to eleven Trotter steps to the get the ground state energy with required precision. This would require more than 800 gates for each of the algorithms.

Ref.~\cite{Seeley2012} reported a strategy which works in bringing down the number of Trotter steps required to achieve a given precision in ground state energy. We refer to it as \emph{magnitude ordering}. For this strategy, the terms containing $\sigma^z$  are separated from terms containing $\sigma^x$ and $\sigma^y$. Then, the terms are arranged in increasing order of the magnitude of their coefficients. The terms containing $\sigma^x$ and $\sigma^y$ are padded in between the terms containing $\sigma^z$. Since the number of $\sigma^x$ and $\sigma^y$ terms are less than $\sigma^z$ terms, the extra $\sigma^z$ terms are applied at the end. The number of Trotter steps required with this strategy are four for Jordan-Wigner and Bravyi-Kitaev. 
\begin{figure}
	\includegraphics[width=9cm]{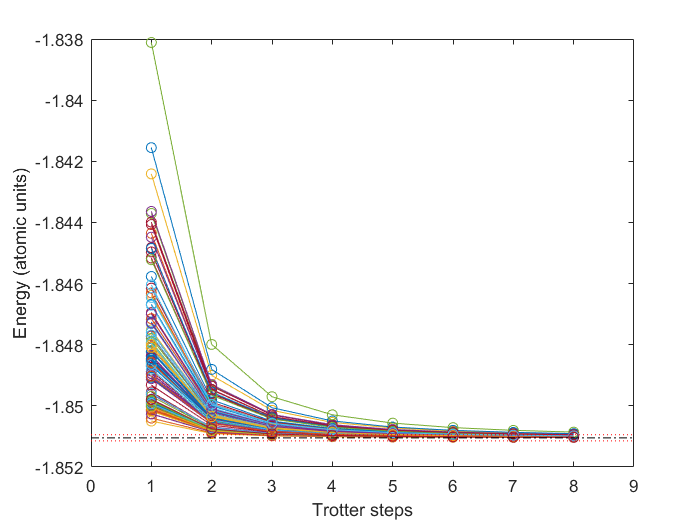}
	\caption{Plots of ground state energies as a function of Trotter steps used to approximate time evolution using the Jordan-Wigner algorithm. Different lines correspond to 1000 orderings of the exponentiated qubit operators of the Jordan-Wigner Hamiltonian.}
    \label{fig:jw_perm}
 \end{figure}
In general, it is not possible to come up with the best ordering because the number of possible orderings grow as a factorial of the number of terms in the Hamiltonian. But, it is possible to generate random permutations to get the behavior of the error. For our analysis we used 1000 randomly generated orderings and ran the code multiple times for all three algorithms. A graph of ground state energy for given number of Trotter steps with 1000 different orderings is plotted in Figure \ref{fig:jw_perm} for Jordan-Wigner. Similar results were found for Bravyi-Kitaev and BKSF algorithms.

\begin{figure}
	\includegraphics[width=9cm]{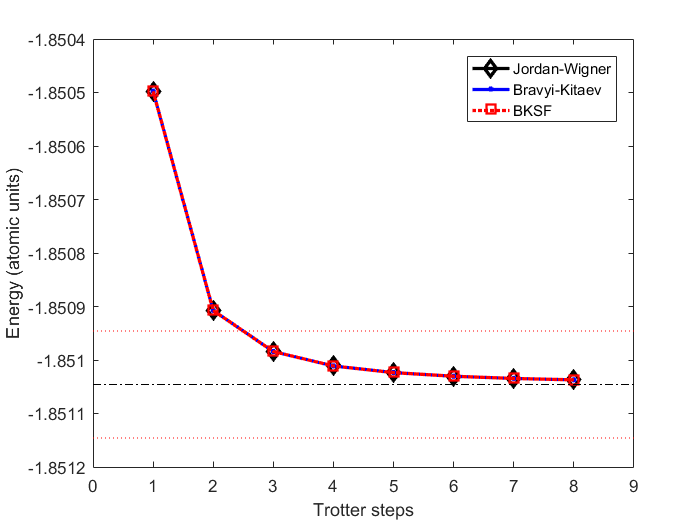}
	\caption{Ground state energy as a function of Trotter steps for the orderings that give the lowest error.  For each of the three methods, we found an ordering that achieved the plotted Trotter error.}
    \label{fig:jw_bk_bksf}
 \end{figure}

We point out that for each ordering of the Jordan-Wigner terms, we were able to find a corresponding ordering for both Bravyi-Kitaev and BKSF that achieved the same error. One of the orderings for BKSF which was found using random permutations and which gives minimum error is given in Eq. \eqref{eq:H_BKSF}. Our results for hydrogen indicate that the Bravyi-Kitaev and Jordan-Wigner spin Hamiltonian terms can be aligned so that the same reordering of the terms yield in the same Trotter error.  This is illustrated in Figure \ref{fig:jw_bk_bksf}, where the Trotter error of all three algorithms collapse onto the same curve.

The improved ordering scheme gave single-step Trotter errors in the ground state energy of $5.4803\cross 10^{-4}$ Hartrees. To obtain errors less than $10^{-4}$ Hartrees, three Trotter steps are required. This translates to a gate requirement of 246, 222, and 237 for Jordan-Wigner, Bravyi-Kitaev and BKSF respectively.  
Further, even in a sample of $1000$ random orderings there were more than one ordering that gave minimum error. 

To get an idea about when BKSF might perform better compared to Jordan-Wigner and Bravyi-Kitaev, we can consider how each of these algorithms scale assuming spatially local interactions. Analyzing the expressions of edge operators, it can be seen that both $\tilde{B}_{i}$ and $\tilde{A}_{ij}$, Eq. \eqref{B_i} and \eqref{A_ij}, scale as $O(d)$, where $d$ is the degree of the graph. Each term in the Hamiltonian in BKSF, scales linearly with the degree of the graph. Whereas each term in the Hamiltonian scales as $O(M)$ and $O(\log(M))$ for Jordan-Wigner and Bravyi-Kiteav algorithm, where $M$ is the number of modes. Now, if we assume that the interaction is reasonably local in a given molecule, that is the degree is bounded, then scaling of BKSF algorithm will be proportional to the number of terms in the Hamiltonian times $O(d)$. Whereas, the scaling of Jordan-Wigner will be proportional to the number of terms in Hamiltonian, times $O(M)$. Since the number of terms in the Hamiltonian will be of the same order for both the cases, BKSF will scale better compared to Jordan-Wigner algorithm asymptotically.

From the results for the hydrogen molecule, it can be seen that the gate count of BKSF is more than Bravyi-Kitaev but less than Jordan-Wigner. The degree of the graph for BKSF algorithm is two and the number of modes are four. 
For small and medium sized molecules, due to the overlap of the atomic orbitals, the  maximum degree of the graph will increase.  This also depends on the extent of the orbitals but as noted above  the Coulomb interactions will not contribute edges to the interaction graph.  Thus, for systems with sufficiently localized orbitals, we expect BKSF to perform better.    

\section{Conclusion}
In this article, we have explored the Bravyi-Kitaev Superfast simulation and illustrated the method on the hydrogen atom.  The implementation of the algorithm can be found online at OpenFermion \cite{Mcclean2017}.  We have compared the Trotter-error for BKSF, Jordan-Wigner and Bravyi-Kitaev algorithm. 

We are currently extending our investigation to larger molecules and observing the importance of choosing the correct one-body basis.  In the present study, we only looked at the hydrogen Hamiltonian in the molecular orbital basis but as the one-body basis changes, the interaction graph also changes.  Other algorithms for time evolution and measurement e.g.~\cite{Childs2015,Mcclean2016,Paesani2017} are compatible with the BKSF scheme outlined here although specifics remain to be worked out.

Also, we have not considered quantum architecture limitations.  OpenFermion can be used to test BKSF and other methods on actual quantum hardware.  It will be particularly interesting to study the noise properties of the BKSF model given its close connection to topological quantum error correction.  This also requires further exploration.

\section*{Acknowledgements}
The authors would like to thank T. Hardikar for helpful comments on the manuscript, and R. Babbush, T. H\"aner, D. Steiger, and J. McClean for their help with OpenFermion.  The authors acknowledge financial support from the Walter and Constance Burke Research Initiation Award.  

\bibliographystyle{apsrev}
\bibliography{Bk2bib}

\end{document}